\documentclass[twocolumn,showpacs,showkeys]{revtex4}
\usepackage{graphicx}
\usepackage{bm}
\usepackage{color}
\usepackage{amsmath}
\usepackage{natbib}

\begin{document}

\title{Radiative corrections to the Coulomb law and model of dense quantum plasmas: Dispersion of waves in magnetized quantum plasmas}

\author{Pavel A. Andreev}
\email{andreevpa@physics.msu.ru}
\affiliation{Faculty of physics, Lomonosov Moscow State University, Moscow, Russian Federation.}

\date{\today}

\begin{abstract}
Two kinds of the quantum electrodynamic radiative corrections to electromagnetic interaction and their influence on properties of highly dense quantum plasmas are considered. Linear radiative correction to the Coulomb interaction are considered. Its contribution in the spectrum of the Langmuir waves is presented. The second kind of the radiative corrections is related to nonlinearity of the Maxwell equations for strong electromagnetic field. Its contribution in spectrum of transverse waves of magnetized plasmas is briefly discussed. At consideration of the Langmuir wave spectrum we included effect of different distribution of the spin-up and spin-down electrons revealing in a shift of the Fermi pressure.
\end{abstract}


\maketitle



\section{Introduction}
Quantum hydrodynamic (QHD) finds applications in area of highly dense astrophysical plasmas: white dwarfs and atmospheres of neutron stars. It happens since more quantum effects display there in compare with semiconductors and metals. High densities lead to small interparticle distances and large wave vectors.

At small interparticle distances and large Fermi velocities the semi-relativistic (weakly relativistic) effects describing by the Breit Hamiltonian give contribution in properties of quantum plasmas. Hence one finds contributions of the quantum Bohm potential, the semi-relativistic correction to the quantum Bohm potential, the relativistic correction to the Fermi pressure, the Darwin interaction, and the semi-relativistic correction to the Coulomb interaction \cite{Shukla PhUsp 2010}, \cite{Shukla RMP 11}, \cite{Ivanov Darwin}. However we can expect the presence of other classic electrodynamic and quantum electrodynamic effects. An example of classic effects is the radiation damping (see Refs. \cite{Landau 2}, \cite{Kuzmenkov 80th}, \cite{Andreev arxiv Rad Damp}). Speaking of quantum electrodynamic effects we mean effects are not presented in the Dirac equation or in the Breit Hamiltonian. One of such effects is the anomalous part of electron magnetic moment. In has been easily included in literature by replacement $\gamma_{0}=\frac{q_{e}\hbar}{2mc}\rightarrow\gamma=g\frac{q_{e}\hbar}{2mc}$, with $g=(1+\alpha/(2\pi))=1.00116$, where $\alpha=e^{2}/(\hbar c)$ is the fine structure constant \cite{Vagin Izv RAN 06}, \cite{Brodin PRL 08}. Another example of the quantum electrodynamic effects is the radiative corrections to the Coulomb interaction \cite{Landau 4}, \cite{Serber PR 35}, \cite{Uehling PR 35}. It appears as a modification of the Coulomb interaction
\begin{equation}\label{QED replacement}\frac{1}{\xi}\rightarrow\frac{1}{\xi}\biggl[1 +\frac{\alpha}{4\sqrt{\pi}}\biggl(\frac{\hbar}{mc\xi}\biggr)^{\frac{3}{2}}\exp\biggl(-\frac{2mc\xi}{\hbar}\biggr)\biggr].\end{equation}
Replacement (\ref{QED replacement}) can be described as the vacuum polarization around point-like charge.

This paper is dedicated to development of the many-particle QHD model \cite{MaksimovTMP 1999}, \cite{Andreev PRB 11} of highly dense plasmas with the radiative corrections to the Coulomb law. We are also going to apply the model to description of waves in magnetized quantum plasmas.

The radiative correction (\ref{QED replacement}) sometimes called linear correction. There are also non-linear radiative corrections revealing in a vacuum polarization $\textbf{P}_{vac}$ and a vacuum magnetization $\textbf{M}_{vac}$, which exist even in absence of medium being non-linear on the electric and magnetic fields \cite{Landau 4}.

The Lagrange function corresponding to the usual Maxwell equations is \cite{Landau 2}, \cite{Landau 4}
\begin{equation}\label{QED L0} L_{0}=\frac{1}{8\pi}(\textbf{E}^{2}-\textbf{B}^{2}).\end{equation}
Small radiative corrections appears as an extra term in the Lagrange function $L=L_{0}+\tilde{L}$ \cite{Landau 4}, where
\begin{equation}\label{QED L extra} \tilde{L}=\frac{\hbar^{2}e^{4}}{45\cdot8\pi^{2}m^{4}c^{6}}[(\textbf{E}^{2}-\textbf{B}^{2})^{2}+7(\textbf{E}\textbf{B})^{2}].\end{equation}

The vacuum polarization $\textbf{P}_{vac}$ and magnetization $\textbf{M}_{vac}$ appears from the Lagrange function as follows $\textbf{P}_{vac}=\frac{\partial \tilde{L}}{\partial \textbf{E}}$, and $\textbf{M}_{vac}=\frac{\partial \tilde{L}}{\partial \textbf{B}}$ \cite{Landau 4}. Nonlinearity of the Maxwell equations described by formula (\ref{QED L extra}) reveals in a distortion of the field of a motionless charge $q$ as \cite{Landau 4}
\begin{equation}\label{QED Coul mod long} \varphi=\frac{q}{r}\biggl(1-\frac{2\alpha^{2}q^{2}\hbar^{4}}{225\pi m^{4}c^{4}r^{4}}\biggr).\end{equation}
Correction in formula (\ref{QED Coul mod long}) has higher order on the fine structure constant $\alpha$ than correction (\ref{QED replacement}), but it slowly decreases with the distance from charge.

However contribution of the vacuum polarization and magnetization gives rather small contribution in the spectrum of the electromagnetic wave. Even in extreme external magnetic field it gives shift of the Langmuir frequency about $10^{-19}\omega_{Le}^{2}$. Thus we are focused on the linear radiative correction (\ref{QED replacement}).

In 2001, after explicit account of the Coulomb and spin-spin interactions in the many-particle QHD including exchange part of these interactions \cite{MaksimovTMP 1999}, \cite{MaksimovTMP 2001}, \cite{MaksimovTMP 2001 b}, the program of including of all semi-relativistic effects in many-particle QHD was formulated \cite{MaksimovTMP 2001}. Most of the semi-relativistic effects are presented by the Breit Hamiltonian \cite{Landau 4}, \cite{MaksimovTMP 2001}, \cite{Ivanov arxiv big 14}. However, there is the annihilation interaction between electrons and positrons \cite{Landau 4}, \cite{Pirenne 1947}-\cite{Andreev arXiv 14 positrons}. All of these interactions were considered in past years \cite{Ivanov Darwin}, \cite{MaksimovTMP 2001}, \cite{Andreev arXiv 14 positrons}-\cite{Andreev IJMP 12}. We are focused now on effects beyond semi-relativistic approach as the vacuum polarization described above. Koide \cite{Koide PRC 13} have recently suggested another method of QHD equation obtaining for the many-particle systems.

Having some recent achievements in classic relativistic hydrodynamics and kinetics \cite{Kuzmenkov 91}-\cite{Mahajan 03} we find their continuous applications \cite{Luca Comisso arXiv 14}, \cite{Asenjo PRE 12}, \cite{Lopez PRE 13}. Steady interest to quantum relativistic effects has been formed in literature \cite{Ivanov Darwin}, \cite{Andreev arXiv 14 positrons}, \cite{Andreev IJMP 12}, \cite{Haas PRE 12}-\cite{Takabayasi PTP 83}. Let us give a brief description of relativistic effects, which have been included in kinetic and hydrodynamic methods of quantum plasma description.
The spin-spin \cite{MaksimovTMP 2001},  the spin-current \cite{Andreev RPJ 07}, the spin-orbit \cite{pavelproc}, \cite{Andreev IJMP 12}, \cite{Asenjo NJP 12}, \cite{Trukhanova EPJD 13}, the current-current \cite{Ivanov Darwin}, \cite{Ivanov arxiv big 14}, \cite{Ivanov RPJ 13} and the Darwin \cite{Ivanov Darwin}, \cite{Asenjo NJP 12} interactions were considered in quantum hydrodynamics and quantum kinetics. Semi-relativistic correction to the kinetic energy, relative force fields and relativistic part of the quantum Bohm potential were considered in Refs. \cite{Ivanov Darwin}, \cite{Ivanov arxiv big 14}, \cite{Ivanov RPJ 13}. Fluidization of the Dirac equation was performed in Refs. \cite{Asenjo PhysPlasm 11}, \cite{Takabayasi PTP 83}. Fluidization of the Klein-Gordon \cite{Haas PRE 12} equation has been done as well. Moreover fluidization of the square-root
Klein-Gordon-Poisson system was considered in Ref. \cite{Haas JPP 13}. The annihilation interaction \cite{Pirenne 1947}, \cite{Berestetski ZETP 49 a}, \cite{Berestetski ZETP 49 b}, which is an interaction specific for electron-positron plasmas, was considered in terms of quantum hydrodynamics and kinetics in Ref. \cite{Andreev arXiv 14 positrons}. It requires consideration of a many-particle formalism, since it leaves no trace in single particle Hamiltonians. Some steps towards semi-relativistic quantum hydrodynamics and kinetics based on the Breit Hamiltonian were also made over 1977-1980 (see Refs. \cite{Goldstein Ph 77}-\cite{Jones PF 80 2}). The Wigner kinetic was applied to the Klein-Gordon equation \cite{Mendonca PP 11} and the Dirac equation \cite{Zhu PPCF 12}, that allowed to get relativistic spectrum of the Langmuir and electromagnetic waves. Relativistic gamma factors were included in the continuity and Euler equations to obtain their relativistic generalization \cite{Akbari-Moghanjoughi PP 13}. Equilibrium relativistic Fermi-Dirac distribution of electrons following motion of cold and non-degenerate ions was considered in Ref. \cite{Shah PP 11} in order to get the nonlinear properties of ion acoustic waves in a relativistic degenerate plasma. Some of mentioned results are discussed in review  \cite{Uzdensky arxiv review 14}.

This paper is organized as follows.  In Sec. II we present the many-particle QHD model for dense quantum plasmas with radiative corrections to the Coulomb law. In Sec III we describe contribution of radiative corrections in dispersion of longitudinal waves in magnetized plasmas. In Sec. IV brief summary of obtained results is presented.

\section{\label{sec:level1} Model}

Rigorous derivation of the many-particle QHD equations for plasmas was performed in 1999 \cite{MaksimovTMP 1999}, whereas its single particle analog was obtained by Madelung in 1926 \cite{Madelung}. The first application of the QHD equations to plasmas was made by Rand in 1964 \cite{Rand PF 64}, where Rand used similarity between quantum one particle hydrodynamics and classic many-particle hydrodynamics. Fluidization of the Pauli equation was done by Takabayasi in 1954-1955 \cite{Takabayasi PTP 55 b}. Application of single particle Hamiltonians has been actively used in past years \cite{Haas PRE 00}-\cite{Brodin NJP 07}. It has been applied along with model kinetic equation \cite{Brodin PRL 08}.

Here, for the first time, we present equations of the many-particle quantum hydrodynamics with the radiative corrections to the Coulomb law. In this section we consider linear the radiative corrections.

The QHD equations appears as
\begin{equation}\label{QED cont eq electrons}
\partial_{t}n+\nabla(n\textbf{v})=0, \end{equation}
and
$$mn(\partial_{t}+\textbf{v}\nabla)\textbf{v}+\nabla p-\frac{\hbar^{2}}{4m}n\nabla\Biggl(\frac{\triangle n}{n}-\frac{(\nabla n)^{2}}{2n^{2}}\Biggr)$$
$$-\frac{\hbar^{4}}{8m^{3}c^{2}}\nabla\triangle\triangle n =q_{e}n\biggl(\textbf{E}_{ext}+\frac{1}{c}[\textbf{v},\textbf{B}_{ext}]\biggr)$$
\begin{equation}\label{QED Euler eq INT}  -q_{e}^{2}n\nabla\int G_{QED}(\mid\textbf{r}-\textbf{r}'\mid)n(\textbf{r}',t)d\textbf{r}',\end{equation}
where
\begin{equation}\label{QED}G_{QED}(\xi)=\frac{1}{\xi}\biggl[1 +\frac{\alpha}{4\sqrt{\pi}}\biggl(\frac{\hbar}{mc\xi}\biggr)^{\frac{3}{2}}\exp\biggl(-\frac{2mc\xi}{\hbar}\biggr)\biggr]\end{equation}
is the Green function of the Coulomb interaction including radiative corrections to the Coulomb law, with $\mbox{\boldmath $\xi$}=\textbf{r}-\textbf{r}'$, $\xi=\mid\mbox{\boldmath $\xi$}\mid$, $\alpha=e^{2}/(\hbar c)=1/137$ is the fine structure constant.

Equation (\ref{QED cont eq electrons}) is the continuity equation containing the particle concentration $n$ and the velocity field $\textbf{v}$. The next equation is the Euler equation describing the momentum balance. The first term on the left-hand side of equation (\ref{QED Euler eq INT}) is the substantial time derivative of the velocity field. The next term is the gradient of thermal pressure related to distribution of particles on different quantum states. The third term, which is proportional $\hbar^{2}$, is the non-relativistic part of the quantum Bohm potential. The last term on the left-hand side is the linear part of the semi-relativistic part of the quantum Bohm potential. Full semi-relativistic quantum Bohm potential can be found in Refs. \cite{Ivanov Darwin}, \cite{Ivanov arxiv big 14}, where the spinless part of the semi-relativistic force field appearing from the Breit Hamiltonian or the Darwin Lagrangian can be found as well.

On the right-hand side of equation (\ref{QED Euler eq INT}) we have two groups of terms. The first of them is the Lorentz force caused by an external electromagnetic field. The second term describes interparticle interaction. It describes the Coulomb interaction modified by the radiative corrections as it is shown by formula (\ref{QED replacement}).

Neglecting the quantum part of the Green function of the Coulomb interaction we can explicitly introduce classic electric field
\begin{equation}\label{QED}\textbf{E}_{int}=-q_{e}\nabla\int G_{0}(\mid\textbf{r}-\textbf{r}'\mid)n_{e}(\textbf{r}',t)d\textbf{r}',\end{equation}
with
\begin{equation}\label{QED}G_{0}(\xi)=\frac{1}{\xi}.\end{equation}
$G_{0}(\xi)$ is the classic part of the Green function of the Coulomb interaction.

In absence of the quantum part of the Green function of the Coulomb interaction we can represent the Euler equation in more familiar form
$$mn(\partial_{t}+\textbf{v}\nabla)\textbf{v}+\nabla p-\frac{\hbar^{2}}{4m}n\nabla\Biggl(\frac{\triangle n}{n}-\frac{(\nabla n)^{2}}{2n^{2}}\Biggr)$$
\begin{equation}\label{QED Euler eq non INT nonRel} -\frac{\hbar^{4}}{8m^{3}c^{2}}\nabla\triangle\triangle n =q_{e}n\biggl(\textbf{E}_{ext}+\textbf{E}_{int}+\frac{1}{c}[\textbf{v},\textbf{B}_{ext}]\biggr),\end{equation}
where
\begin{equation}\label{QED div E usual} \mathrm{div}\textbf{E}_{int}=4\pi\rho,\end{equation}
and
\begin{equation}\label{QED curl E usual} \textrm{curl}\textbf{E}_{int}=0,\end{equation}
with the charge density $\rho=\sum_{a=e,i}(q_{a}n_{a})$.

In equations (\ref{QED Euler eq non INT nonRel})-(\ref{QED curl E usual}) the Coulomb interaction is presented in traditional non-integral form in terms of the electric field $\textbf{E}_{int}$ obeying the quasi-electrostatic Maxwell equations (\ref{QED div E usual}), (\ref{QED curl E usual}). Equation (\ref{QED Euler eq non INT nonRel}) shows relation of equation (\ref{QED Euler eq INT}) to traditional form of hydrodynamic equations.

\section{Dispersion of high frequency longitudinal waves in magnetized plasmas}

In this section we consider small perturbations of equilibrium state describing by nonzero particle concentration $n_{0}$, and zero velocity field $\textbf{v}_{0}=0$ and electric field $\textbf{E}_{0}=0$.

Assuming that perturbations are monochromatic
\begin{equation}\label{EXCH_SEC perturbations}
\left(\begin{array}{ccc} \delta n
 \\
\delta \textbf{v}
 \\
\end{array}\right)=
\left(\begin{array}{ccc}
N_{A} \\
\textbf{V}_{A} \\
\end{array}\right)e^{-\imath\omega t+\imath \textbf{k} \textbf{r}},\end{equation}
we get a set of linear algebraic equations relatively to $N_{A}$ and $V_{A}$. Condition of existence of nonzero solutions for amplitudes of perturbations gives us a dispersion equation.

We now get the linearized set of QHD equations (\ref{QED cont eq electrons}) and (\ref{QED Euler eq INT})
\begin{equation}\label{QED cont eq electrons LIN}
\partial_{t}\delta n+n_{0}\nabla\delta\textbf{v}=0, \end{equation}
and
$$mn_{0}\partial_{t}\delta\textbf{v}+\frac{\partial p}{\partial n}\nabla \delta n-\frac{\hbar^{2}}{4m}\nabla\triangle \delta n$$
$$-\frac{\hbar^{4}}{8m^{3}c^{2}}\nabla\triangle\triangle \delta n=q_{e}n_{0}\frac{1}{c}[\delta\textbf{v},\textbf{B}_{ext}]$$
$$-q_{e}^{2}n_{0}\Biggl[\frac{4\pi}{k^{2}} - \sqrt{2\pi}\frac{e^{2}}{mc^{2}}\frac{4\pi}{k}\times$$
\begin{equation}\label{QED Euler eq LIN}   \times\sqrt[4]{\frac{1}{4}\biggl(\frac{\hbar}{mc}\biggr)^{2}k^{2}+1}\cdot\sin\biggl(\frac{1}{2}\textrm{arctg}\frac{\hbar k}{2mc}\biggr)\Biggr]\nabla\delta n.\end{equation}

We have linearized form of QHD equations, where interaction is presented by integral terms. This is why we do not see familiar $q_{e}n_{0}\delta \textbf{E}$ on the right-hand side of the linearized Euler equation (\ref{QED Euler eq LIN}). Instead of it we have the first term in the large square brackets. Whereas the second term in the brackets is the radiative corrections of the Coulomb interaction.

We apply the following equation of state to get closed set of QHD equations
$$p_{sf}=\frac{(6\pi^{2})^{\frac{2}{3}}}{10}\frac{\hbar^{2}}{m}\times$$
$$\times\Biggl[\biggl(\frac{n_{0}}{2}+\frac{\Delta n}{2}\biggr)^{\frac{5}{3}}\biggl[1-\frac{1}{14}\frac{(3\pi^{2})^{\frac{2}{3}}\hbar^{2}}{m^{2}c^{2}}\biggl(\frac{n_{0}}{2}+\frac{\Delta n}{2}\biggr)^{\frac{2}{3}}\biggr]$$
\begin{equation}\label{QED eq State single Fl polarised} +\biggl(\frac{n_{0}}{2}-\frac{\Delta n}{2}\biggr)^{\frac{5}{3}}\biggl[1-\frac{1}{14}\frac{(3\pi^{2})^{\frac{2}{3}}\hbar^{2}}{m^{2}c^{2}}\biggl(\frac{n_{0}}{2}-\frac{\Delta n}{2}\biggr)^{\frac{2}{3}}\biggr]\Biggr],\end{equation} 
where $\Delta n=n_{0}\tanh\biggl(\frac{\gamma_{e}B_{0}}{T_{Fe}}\biggr)$, $\gamma_{e}$ is the gyromagnetic ratio of electron, $B_{0}$ is an external magnetic field, $T_{Fe}$ is the Fermi temperature in units of energy \cite{MaksimovTMP 2001 b}, \cite{Andreev arxiv Two Fl el}. Formula (\ref{QED eq State single Fl polarised}) includes different occupation of quantum states by the spin-up and spin-down electrons, since they have different energy in the external magnetic field.
It corresponds the non-relativistic Fermi pressure $p_{Fe}=\frac{\hbar^{2}}{5m}(3\pi^{2})^{\frac{2}{3}}n^{\frac{5}{3}}$ at low densities and zero external magnetic field.

As the first application we coincide quantum plasmas without external magnetic field and derive spectrum of the Langmuir waves.

Equations (\ref{QED cont eq electrons LIN}) and (\ref{QED Euler eq LIN}) give the following $\omega(k)$ for the Langmuir wave
$$\omega^{2}=\omega_{Le}^{2}\biggl[1- \sqrt{2\pi}k\frac{e^{2}}{mc^{2}}\times$$
$$\times\sqrt[4]{\frac{1}{4}\biggl(\frac{\hbar}{mc}\biggr)^{2}k^{2}+1}\cdot\sin\biggl(\frac{1}{2}\textrm{arctg}\frac{\hbar k}{2mc}\biggr)\biggr]$$
\begin{equation}\label{QED disp Lang 3D} +\frac{v_{Fe}^{2}}{3}\biggl(1-\frac{1}{10}\frac{v_{Fe}^{2}}{c^{2}}\biggr)k^{2}
+\frac{\hbar^{2}k^{4}}{4m^{2}}-\frac{\hbar^{4}k^{6}}{8m^{4}c^{2}} , \end{equation}
where
\begin{equation}\label{QED Langmuir freq 3D} \omega_{Le}^{2}=\frac{4\pi e^2 n_{0}}{m}\end{equation}
is the Langmuir frequency, and
\begin{equation}\label{QED} v_{Fe}=\frac{(3\pi^{2})^{\frac{1}{3}}\hbar n_{0}^{\frac{1}{3}}}{m}\end{equation}
is the Fermi velocity.

Since we do not apply the external magnetic field both terms in equation of state (\ref{QED eq State single Fl polarised}) are equal. This term in $\omega(k)$ coincides with the result of Ref. \cite{Ivanov Darwin}. Our new result is in the modification of the Langmuir frequency caused by the radiative correction.

Main factor showing contribution of the radiative correction to the Coulomb interaction is the ratio of potential energy at wavelength distance to the electron rest mass $\frac{e^{2}k}{mc^{2}}$.

\begin{figure}
\includegraphics[width=8cm,angle=0]{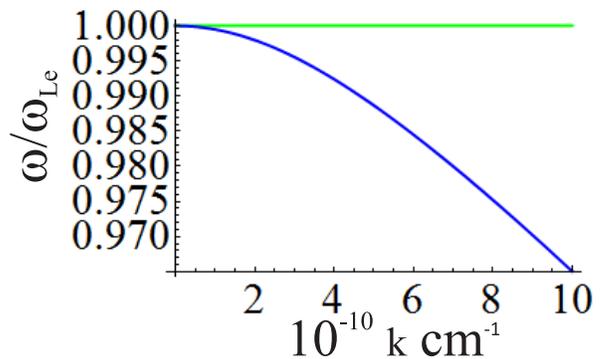}
\caption{\label{QED Fig_RadCorrToLanm} (Color online) The figure shows
effective decreasing of the Laggmuir frequency in the spectrum of the Langmuir wave due to the radiative correction of Coulomb interaction, which is presented by the first group of terms in formula (\ref{QED disp Lang 3D}). Horizontal line presents unchanged Langmuir frequency. The curve line shows the decreasing of the Laggmuir frequency contribution.}
\end{figure}

On Fig. (\ref{QED Fig_RadCorrToLanm}) we present an estimation of the shift of the Langmuir frequency caused by the radiative correction.

In Ref. \cite{Ivanov Darwin} it was found
$$\omega^{2}=\omega_{Le}^{2}\biggl[1-2\frac{v_{B}^{2}}{c^{2}}\biggr]$$
\begin{equation}\label{QED general form of disp eq from Ivanov} +\frac{1}{3}v_{Fe}^{2}\biggl(1-\frac{1}{10}\frac{v_{Fe}^{2}}{c^{2}}\biggr)k^{2} +v_{B}^{2}k^{2}\biggl(1-2\frac{v_{B}^{2}}{c^{2}}\biggr), \end{equation}
with $v_{B}^{2}\equiv\hbar^{2}k^{2}/(4m^{2})$ is the square of Bohm velocity, see also Refs. \cite{Ivanov arxiv big 14}, \cite{Ivanov RPJ 13} for more discussion of the spinless semi-relativistic quantum hydrodynamics.

Formula (\ref{QED general form of disp eq from Ivanov}) does not include the radiative corrections, but it includes the Darwin interaction and the force field existing due to simultaneous account of the semi-relativistic correction to kinetic energy and the Coulomb interaction in the Schrodinger equation.

Comparing formula (\ref{QED disp Lang 3D}) with results of Refs. \cite{Uzdensky arxiv review 14}, \cite{Zhu PPCF 12}, \cite{Mendonca PP 11} we see that simultaneous application of Wigner function and the Dirac and Klein-Gordon equations for getting of quantum kinetics and spectrum of waves in plasmas does not lead to account of the radiative corrections.

\subsection{Spectrum of longitudinal waves propagating in plasmas being in an external magnetic field}

For classic plasmas being in an external magnetic field, the external magnetic field does not change the spectrum of the Langmuir waves propagating parallel to the external magnetic field. In quantum plasmas it is correct if we do not account spin effect coming via the change of the equation of state (\ref{QED eq State single Fl polarised}). It gives the following contribution in the square of the Langmuir wave frequency
\begin{equation}\label{QED } \Delta\omega^{2}_{spin}=-\frac{1}{3}v_{Fe}^{2}\cdot\frac{1}{9}\tanh^{2}\biggl(\frac{\gamma_{e}B_{0}}{T_{Fe}}\biggr)k^{2}\end{equation}
at small spin polarization.

At propagation of the Langmuir waves perpendicular to external magnetic field the square of the cyclotron frequency shifts the square of the Langmuir wave frequency $\Delta\omega^{2}(k)=\Omega^{2}$, where $\Omega=\frac{qB_{0}}{mc}$ is the cyclotron frequency.

\section{Conclusions}

We have derived the set of many-particle QHD equations containing the radiative correction to the Coulomb interaction. The radiation correction reveals itself in the Euler equation in the form of a potential force field. So, it gives contribution in longitudinal plasma waves only. We have considered contribution of the radiative correction in the spectrum of Langmuir waves. We have found that the radiation correction to the Coulomb interaction leads to decreasing of frequency. Magnitude of the decreasing is proportional to the ratio of potential energy of two electron Coulomb interaction to the rest energy of the electron.

We have considered equation of state for semi-relativistic quantum plasmas being in an external magnetic field at zero temperature. Even without consideration of spin evolution, spins of electrons reveals in the equation of state. Spin-up and spin-down electrons occupy quantum states differently. This effects gives contribution in the spectrum of Langmuir waves in magnetized plasmas. This effect appears to be isotropic and gives equal contribution in the spectrum at wave propagation parallel and perpendicular to the external magnetic field.


\begin{acknowledgements}
The author thanks Professor L. S. Kuz'menkov for fruitful discussions.
\end{acknowledgements}

\end{document}